# A Framework for Teaching the Fundamentals of Additive Manufacturing and Enabling Rapid Innovation

**Jamison Go, A. John Hart**
*Department of Mechanical Engineering and Laboratory for Manufacturing and Productivity, Massachusetts Institute of Technology, Cambridge, MA*
Corresponding author: ajhart@mit.edu

# Abstract

The importance of additive manufacturing (AM) to the future of product design and manufacturing systems demands educational programs tailored to embrace its fundamental principles and its innovative potential. Moreover, the breadth and depth of AM spans several traditional disciplines, presenting a challenge to instructors yet giving the opportunity to integrate knowledge via creative and challenging projects. This paper presents our approach to teaching AM at the graduate and advanced undergraduate level, in the form of a 15-week course developed and taught at the Massachusetts Institute of Technology. The lectures begin with in-depth technical analysis of the major AM processes, then focus on special topics including design methods, machine controls, and applications of AM toward current industry needs and emerging materials and length scales. In lab sessions, students operate and characterize desktop AM machines, and work in teams to design and fabricate a bridge having maximum strength per unit weight while conforming to geometric constraints. The class culminates in a semester-long team design-build project; in a single semester of the course, teams created prototype machines for printing molten glass, soft-serve ice cream, biodegradable (chitosan) material, carbon fiber composites, performing large-area parallel extrusion, and for in situ optical scanning during printing. Several of these projects led to patents, future research, and peer-reviewed publications. We conclude that AM education, while arguably rooted in mechanical engineering, must be truly multidisciplinary, and education programs must embrace this context. We also comment on our experiences adapting this course to a manufacturing-focused master's degree program and a one-week professional short program.

# 1. Introduction

Additive Manufacturing (AM), which collectively refers to methods of bottom-up object creation by layer-by-layer material joining [1], expands the design space for complex objects, enables exploration of new material properties, and may change how we configure both local and global supply chains. These and other potential benefits have led to accelerated investment in the validation of current AM technologies, and the rapid exploration of new concepts. As one point of evidence, the worldwide revenue from AM products and services increased by 35.2% from 2013 to 2014, now totaling $4.1 billion [2]. For at least the next several years, it is likely this growth rate will be maintained, and perhaps will increase. Moreover, the influence of the 'maker' movement is marked by an astonishing 346% average annual increase in sales of consumer 3D printers (less than $5000 each) from 2008 to 2011 [3].

Several AM public and private initiatives, including research centers and consultancies, have been established to explore opportunities for innovation and adaptation of AM. A prime example is the National Additive Manufacturing Innovation Institute (NAMII), now called America Makes. This is a collaboration of more than 50 companies, research universities, community colleges, and non-profit organizations to foster the development of emerging technologies in AM [4]. The Nanyang Technological University (NTU) Additive Manufacturing Centre in Singapore hosts over $30 million in state-of-the-art AM machines [5]. The Centre for Doctoral Training (University of Nottingham, Loughborough University, Newcastle University, and University of Liverpool) and Engineering and Physical Sciences Research Council (EPSRC) have a large research team and a four-year PhD training program to produce research and industry leaders in AM [6]. University of Texas El Paso hosts the W.M. Keck Center for 3D Innovation, a 13,000 square foot facility with over 40 AM machines that supports AM research, innovation, and education [7].

The rapid growth and disruptive potential of AM technologies demands education programs that address the fundamental principles of AM and likewise enable designers and engineers to realize its capabilities. Moreover, given the hands-on nature of AM and growing access to AM hardware, software, and materials, AM can be used as a teaching tool in several disciplines. As early as 2009, it was identified that AM education is critical to the advancement of the field, and that programs at the university, industry, technical college, management, and general population levels be implemented [8]. Since then, many new initiatives for AM education have been launched; a few examples are described below.

- The Laboratory School for Advanced Manufacturing, a collaboration between the University of Virginia (Curry School of Education, School of Engineering and Applied Science) , Charlottesville City Schools, and the Albemarle County Public Schools, was established to develop an open-source curriculum for teaching digital fabrication to K-12 students which includes AM among its core fabrication techniques [9].
- Instructors at Virginia Polytechnic Institute and State University (Virginia Tech) and the University of Texas, Austin, report teaching AM design courses as part of their engineering curriculum [10].
- Missouri University of Science and Technology describes the integration of AM into their product realization curriculum. They further describe undergraduate research opportunities, interactions with industry and the results of a community workshop where members were introduced to AM and engaged in the design and printing a pen holder [11].

- Tsinghua University in Beijing, China, reports on several AM-focused educational opportunities for its students, adjacent colleges, and members of industry in multiple disciplines. This includes a description of a four-year undergraduate program in AM and an overview of their research efforts in AM [12].
- The Society of Manufacturing Engineers (SME) in collaboration with the Milwaukee School of Engineering and America Makes have developed an AM certificate program and exam which builds off fundamental manufacturing knowledge [13].
- Utrecht University in the Netherlands offers a 5-day professional course that provides the basics of 3D printing followed by specific challenges with regards to AM and biofabrication [14].
- America Makes hosted a two-day course for engineers, technicians, and managers to learn the essentials of AM from basic processes/limitations to time/cost analysis in the scope of conventional manufacturing [15].
- Stratasys provides a downloadable 3D printing curriculum complete with STL files [16]. This is tailored to general understanding and practice of AM technology, such as for high school level audiences.
- In 2014, Deloitte University offered "3D Opportunity: The course on additive manufacturing for business leaders". This is a brief, digitally delivered course summarizing the implications of AM and highlights commercial opportunities. [17].

The content spectrum in these courses can be traced to the audience and the format. For example, universities have the opportunity to augment core design and manufacturing courses with AM technologies in prototyping as well as create semester-long courses focused on AM. For industry audiences, interest is often greater in the method and applications of AM than the fundamental process science, and short course formats are more appropriate. As a result, several short programs for industry audiences have emerged, and some large corporations are building their own training programs for AM. For all audiences, hands-on exposure to AM machines and/or parts is vital to illustrate the workflow involved in AM, the unique capabilities, and important limitations of exemplary AM techniques.

At MIT we have sought to explore AM education as both a means to enhance understanding and use of AM technology, as well as to teach the underlying principles of machine design, materials processing, and manufacturing operations. Currently, the MIT core courses in manufacturing can only include 1-2 lectures on the general principles of AM [18], along with incorporating AM as a tool to prototype designs in semester projects. This paper presents our approach to teaching AM at the graduate and advanced undergraduate level, in the form of a 15-week course dedicated to the fundamentals and applications of AM. The course includes lectures, laboratory sessions, and hands-on projects. Our experience teaching the course, and the impressive outcomes achieved by the student projects, demonstrate the innovative potential of AM. The experience also highlights the benefits of a multidisciplinary approach to educate and stimulate improvements to AM technology.

## 2. Course Objectives and Structure

The MIT AM course was designed for graduate students and advanced undergraduates. Among graduate students, the course was open to students in all fields. For undergraduates, we sought students who had completed at least one course in mechanical design and/or manufacturing processes. The content was structured to meet following learning objectives:

- To learn the fundamentals of AM of polymers, metals, ceramics, and emerging materials, along with the operating principles of related mainstream AM techniques.
- To gain hands-on experience with desktop AM machines and understand the complete AM sequence by designing, fabricating, and measuring parts and evaluating their performance against functional constraints.
- To learn how to design components for successful use of AM, and to measure parts using three-dimensional scanning techniques.
- To compare and contrast AM processes with conventional manufacturing methods in terms of rate, quality, cost, and flexibility.
- To realize emerging applications of AM across major industries, including medical devices, aerospace and vehicle structures, energy, and consumer products.
- To integrate the above objectives in a cumulative project that proposes, prototypes, and evaluates a novel AM capability.

To implement these objectives, we combined lecture sessions with a sequence of laboratory activities, and hands-on assignments. The sequence of lecture and laboratory topics is shown in Figure 1. In order to build a strong foundation for the project, the first five weeks of the class first presented an overview of the AM industry and technology landscape, and in-depth lectures for each major AM process in parallel with laboratory exercises. This enabled the students to realize the capabilities and limitations of current AM technology early in the semester, and foster appreciation for emerging topics addressed in later lectures.

## 3. Course Content and Outcomes

### 3.1 Lectures and Labs

The beginning of the course is devoted to teaching the fundamentals of additive manufacturing through focus on the cornerstone AM processes: fused deposition modeling (FDM), stereolithography (SLA), and selective laser sintering/melting (SLS/SLM). Each process is addressed in a two-lecture sequence, beginning with the fundamental mechanism of operation, main applications, and history of the technology. Students are guided to derive quantitative relationships among process parameters, build rate, and quality, and to compare these to observations from example components, along with commercial and academic literature.

In parallel with the lectures on FDM and SLA, laboratory sessions introduce students to the corresponding desktop AM machines. Our basic lab workbench, which is typically shared by 2-3 students, is shown in Figure 2. UP! Mini ($599) and Formlabs Form-1+ ($3299) printers are used for FDM and SLA lab sessions, respectively [19], [20]. Each station also includes a 3D scanner, USB microscope (Celestron 44302), and digital calipers used for a metrology lab session [21]. Rather than have the same scanner at each station we chose to present a variety of options and allow the students to evaluate their performance. In 2014, we chose the Next Engine scanner, the Makerbot Digitizer, the Cubify Sense, and the Xbox Kinect with Skanect software [22]–[25]. In 2015, we added the Asus Xtion PRO LIVE with Skanect, Matter and Form, and Fuel 3D Scanify [26]–[28].

In each lab session, teams of students complete an outlined exercise involving part pre-processing using the machine software, printing (while discussing and observing the machine operation), post-processing, and inspection. In the FDM lab, students discuss the gantry-style machine design, study the structural and aesthetic influences of part orientation, observe the placement of support material, and discuss the importance of heating of the print platform and the influence of ambient temperature control. In the SLA lab, students inspect the galvanometer mechanism used to position the laser (from a disassembled machine), discuss the method of resin recoating, learn how post-build curing of parts influences mechanical properties, and compare and contrast process capabilities and support structure designs between FDM and SLA. Through these activities, we aim for the students to learn how to select the appropriate process based on its capabilities (e.g., rate, resolution, and surface quality) and their design goals, and to relate the process capabilities to the method of material deposition and the machine architecture.

Insights from the operation and output of desktop AM machines are complemented by presentation of a collection of parts made using commercial AM machines, including those by Stratasys (Fortus FDM), EOS (polymer SLS and metal SLM), and Arcam (EBM). We find this is a critical instructional element given the growing breadth of industrial AM processes. To enable direct comparisons, we procured the same parts printed on these professional systems as made in lab, along with specifically suited examples from each professional system (e.g., a hip implant cup from Arcam). We also discuss in detail, via videos and schematics, the key machine design principles and components that differentiate industrial AM systems from desktop AM systems.

To teach scanning techniques, we likewise use both lecture and lab sessions. The lecture describes the fundamentals of 3D surface metrology techniques, including surface and volume approaches including contact probing, optical range finding, projected texture, and computed tomography. The performance and limitations of each method are emphasized both qualitatively and quantitatively. In the lab, students are given the opportunity to explore and evaluate each 3D scanner by measuring a variety of objects that vary in geometry, features, and optical properties. Student groups are encouraged to share their observations and results in a post-lab discussion between each scanning lab session. An overarching conclusion is that present-day desktop scanning tools are largely inadequate to capture the detail needed to perform accurate quantitative metrology on AM parts, and it is challenging to process the scan data to enable comparison with the original solid model or to enable use of AM to replicate objects with sub-millimeter detail and accuracy. This is because the majority of desktop scanning tools use optical range-finding methods, which are sensitive to surface finishes and deep part features, as shown in Figure 3. In light of this we showed how surface coatings such as talcum powder can be used to increase scan fidelity with minimal surface roughness defects.

*3.2 Special Topics Lectures*

While the main focus of the second half of the term is the AM innovation project, the lecture sessions continue and focus on more specialized topics. Due to limited time, it is not possible to capture all such topics; however, our approach is to select a series of topics that represents the breadth of interest and background among the students, and to support the project topics where possible. The order of special topics lectures was chosen to transition from the process driven lectures towards emerging developments in AM and finally industry relevant knowledge such as design, economics, and application-oriented case studies. Guest lecturers from industry and academia were invited to speak on their experiences and expertise. For example, Prof. Wojciech Matusik of the MIT Electrical Engineering and Computer Science department gave a lecture on his work in computational design for AM. Maxim Lobovsky, co-founder of Formlabs, spoke about the experience of building Formlabs products and the opportunities for entrepreneurship in AM. We find that the students highly appreciate the guests' perspectives in contrast to the mainstream lecture series.

These topics will certainly evolve with each offering of the course. In previous terms, we have highlighted the following topics (listed in alphabetical order):

- Design for Additive Manufacturing (DFAM): In addition to discussing design rules and limitations for each AM technique during the core lecture series, we discuss formal methods to evaluate the dimensional and feature space capability of AM processes. These include the use of test artifacts, and the classification of features based on their orientation and complexity.

- Machine Controls: A lecture was dedicated to the dynamics of gantry-based motion systems, and scanning mirror systems, to show how these aspects can limit process rate and influence accuracy. Analytical representations for simplified dynamics were derived, and control strategies were discussed and evaluated using numerical simulations of the dynamics.

- Bioprinting: Use of AM with biological and/or biocompatible materials to build tissue scaffolds and novel devices for in vitro laboratory use (e.g., 3D cell structures or organ-on-chip models for drug testing). We also critically evaluate the potential to use AM to fabricate replacement parts for living tissues and organs in the future.

- Printed Electronics: We address the combination of AM and integrated circuit assembly methods (e.g. pick-and-place) to fabricate hybrid structural/electronic parts, along with current and emerging methods for direct deposition of electronic materials such as conductive and semiconducting inks.

- Digital Assembly: Contrasting continuous material deposition schemes, and building upon electronics pick-and-place methods, we discuss use of discrete assembly as the AM process itself. Academic demonstrations for voxellated materials and cellular composites are reviewed, and scalability of the approach is analyzed and discussed.

- Micro- and Nanoscale AM: We discuss the scalability of AM techniques to miniature dimensions, the properties and applications of AM at these scales, and the related challenges and opportunities in material and process development. Our discussion included micro-scale of functional inks, micro-SLA and two-photon nanoscale patterning, and micro-SLS of metal powders.

- Computational Design: We reviewed leading techniques processing 3D geometries, and for distribution of material specifications (e.g., porosity, stiffness), though it was beyond scope to discuss the underlying algorithms.

- Economics of AM: We discussed studies evaluating the cost of AM in several exemplary cases, and how these costs vary with process type, part complexity, and production volume. The economics of AM were compared to traditional manufacturing processes such as machining and injection molding.

- Entrepreneurship in AM: Invited guests from local startups, and representatives from the university technology licensing office described their experiences along with the intellectual property landscape for AM technology.

*3.3 Individual Printing Assignment*

Following the FDM and SLA labs, we asked each student to design a part, or an assembly of parts, with aim to investigate the limiting attributes of a chosen AM process or combination of processes. Students were not restricted in their designs or choice of machines; they could use any machine to which they had access, including in the teaching lab. The only consideration asked of students was that their part(s) should be difficult to manufacture using traditional processes and/or be suited to customization. Before printing their parts, the students were asked to consider how the chosen process would influence the aesthetics and functionality of their part(s). During the print, students were asked to observe the fabrication of key features so they could relate the process operation to defects observed later.

Nearly all of the students used the FDM and SLA machines in the teaching lab, and a collection of objects is shown in Figure 4. Other students used comparable desktop FDM and SLA systems in their own labs, and some used professional FDM systems. The most common objects included personalized accessories for electronics, small figurines, or jewelry; for example, one student designed a cable organizer for his phone charger while another designed a custom case for a circuit board she had made.

After printing and inspecting their parts, the students summarized their findings in a report, including the following questions:
- What part(s) did you print and why is this best manufactured additively?
- Which printer did you use and how did you expect its limitations would affect the outcome of the part?
- Where can defects be found observed and why do these occur using knowledge about the printing process?
- Which features of the part are critical to its functionality and how do their measured values compared to the values specified in design? Speculate on these differences using knowledge of the printing process and machine capabilities.
- How could features of the part be redesigned to ensure better quality or functionality?
- How could the process be modified to reduce or remove observed defects?
- Could a better result be expected using another AM process, and if so why?

Major emphasis was placed on the identification and analysis of defects, i.e., non-idealities in the part geometry as well as imperfections in the microstructure. Example defects noted by students are shown in Figure 5. In 5a, the student noted significant surface roughness along a vertically oriented edge of an FDM part, which was attributed to the layer-by-layer process of FDM as the bumps were periodic and oriented along the Z-axis of the printer. In Figure 5b, the student also identifies the discretization of circular features due to the layer orientation. In Figure 5c, we see how sharp corners in the CAD model are rounded in the part, which is attributed to the circular cross section of the extrusion orifice and sometimes the overflow of material as the FDM extruder turns the corner. This is a common feature of parts printed using desktop FDM machines, because the extrusion rate is constant while the gantry decelerates and accelerates to change direction, resulting in excess material at corners. Another student who printed on an SLA machine discussed dimples on the surface of his part (Figure 5d), remaining from break-away of the support structures. The student further noted that light sanding after cleaning and drying the part was effective in removing these bumps.

The students were also asked to measure key dimensions (e.g., height, width, and hole diameter) of their parts, and compare these to the nominal expected dimensions from their CAD models. Students were permitted to use any metrology equipment that was suitable for the task; at minimum, calipers, USB microscopes, and the 3D scanners mentioned above were available in the lab. Students measured each key dimension five times and reported the statistics (mean, standard deviation) relative to the nominal value and the capability of the measurement tool (quoted accuracy and repeatability). Overall, students noticed that printed parts had larger exterior dimensions (especially where support material was used) than nominal, smaller interior dimensions nominal, and accuracy was greatest for features oriented in the XY plane. This provided a further opportunity for the students to realize how the material deposition fundamentals and machine performance affect the quality of components made by AM.

*3.4 Bridge Challenge*

After the students became familiar with the operation and limitations of basic AM technologies via the labs and printing assignments, we sought to present an open-ended challenge that would embrace the design freedom of AM. Inspired by bridge building competitions that are often found in civil engineering curricula, we asked the students to work in teams (3-4 students per team) to design and construct a small bridge having maximum specific strength. The bridges were to be made only using the desktop FDM and/or SLA machines in the lab, with stock materials (ABS filament and clear Formlabs resin), and assembled without the use of adhesives or mechanical fasteners. Each bridge was required to span a 12" gap and have no larger than a 4" x 4" x 12" volume (before loading). For testing, each bridge was mounted onto a custom-built rig (Figure 6) and loaded using a modified Instron machine (Model 1122; equipped with a 1 kN or 10 kN load cell based on the expected maximum load. The winning bridge was determined by measuring the maximum load before failure, divided by the bridge weight.

Most students in the course had at least an introductory background in solid mechanics, and the AM bridge challenge required them to combine this knowledge with their awareness of how AM part design, process, and microstructure influences mechanics. To apply quantitative analysis to their bridge designs, it was therefore necessary for students to perform experiments to assess the mechanics of AM parts and consider practical issues such as anisotropy, voids and surface defects, and the influence of post-cure conditions. Also, the use of the teaching lab printers was an important constraint because the specified span length was more than twice the longest printing dimension of the lab printers. As a result, many bridges were made from joining multiple parts.

Selected bridge designs and design methodologies are shown in Figure 7. The bridge shown in Figure 7a was designed using topology optimization tools; the team began with the stated volume constraint, defined the mounting and load points, and generated an optimized geometry based on this information. The bridge pictured in Figure 7b was inspired by the bone structures within bird wings. Most teams utilized finite element analysis tools to simulate their bridges under loading, such as the team that designed the bridge in Figure 7c.

The winning bridge design shown in Figure 8 demonstrated remarkable creativity and understanding of AM concepts. The team used an FDM machine to print a flexible ABS ribbon with a thickness of only one extrusion 'road'. Comprised of a single-piece ribbon, the bridge was incredibly lightweight but strong because the printed orientation of the ribbon put full lengths of printed ABS in tension. The single strand thickness also allowed the structure to be flexible, enabling the bridge to be printed as a single piece in a single job and unravel to span the gap. To further leverage the capabilities of AM, the bridge incorporated monolithic flexible hooks to latch onto the aluminum beams of test rig. This design weighed 24 grams and took 3 hours to print yet withstood 1614 N of force at the loading point.

On the bridge competition day, the students briefly presented their design, methodology, analysis techniques, and construction. This presentation was given while their bridge was being setup on the test rig. The test followed soon after, and high-speed video was captured and replayed shortly after the test.

*3.5 AM Innovation Project*

The capstone of the course is a project whereby students are asked to conceptualize, justify, design, and prototype a novel contribution to AM technology. When selecting a project topic, teams are asked to draw from technical and practical motives discussed in the introductory lectures and combine these motives with their own interests. Because AM is inherently multidisciplinary, we do not impose any limitations on the projects other than feasibility for completion within the semester, and suitability of the level of resources ($1000 budget and machine shop access). To stimulate brainstorming a list of thematic challenges was given to the students, including printing of novel materials (e.g., food, composites, biomaterials), making improvements in resolution or quality of a specific AM process, and building multi-mode AM equipment (e.g., combining different deposition methods, or deposition and measurement).

Considering the ambitious objectives of the project, we sought to engage the students toward this goal as early as possible. During the first week, we asked the students form project teams based on common interests, and to brainstorm project concepts as they complete the introductory lectures and laboratories during the first four weeks. After that, a series of milestones was fulfilled. The first milestone was a project proposal in which teams: described their project concept and its importance to AM; reviewed academic, patent, and commercial literature; set rough goals for the end of the semester; and identified the skills and resources needed to successfully complete the project. The second milestone was a midterm project review presentation where teams delivered a refined project proposal, along with initial analysis, proof-of-concept experiments, and a concrete set of objectives for the final prototype. For the reminder of the semester, the teams met with the instructional staff at least on a weekly basis to review their progress and receive feedback.

Because many of the project concepts were considered to be novel, we facilitated meetings between the teams and the MIT Technology Licensing Office (TLO) such that provisional patents could be filed when appropriate, prior to the final expo. This step was crucial as many of the projects stimulated great interest and teams were contacted later for detailed public disclosure. The seven final projects completed in the Spring 2014 semester are summarized below; a select few are then described in further detail.

- Fused deposition modeling of soft serve ice cream (Figure 9) – Students designed and built a system for extrusion-based AM of soft-serve ice cream. A commercially available soft serve machine was combined with a 3-axis positioning gantry and liquid nitrogen rapid cooling system to create edible 3D structures from ice cream.
- Gravity-driven 3D printing of optically transparent glass (Figure 10) – Students created the first-ever system for AM of structures directly from molten glass. The printer featured a heated kiln (into which molten glass was loaded) atop a X-Y gantry. 3D structures were created using gravity-fed deposition from the kiln and controlled XY positioning. The platform moved downward after each layer was completed.
- High-throughput printing of polymers using local surface tension manipulation (Figure 11) – Students realized a concept for simultaneous deposition of polymer droplets by heating digitally indexed areas of a mesh wire grid equal in size to the build platform. Polymer in these areas melted and formed droplets that were transferred to the build platform. The concept was demonstrated using a benchtop rig with a small wire grid, and analysis of the thermal and fluid dynamics of the process was performed.
- Rapid prototyping with biodegradable composites – Students built a syringe-based multi-material extrusion system with a mixing nozzle to explore the AM of biocompatible composite parts using hydrogels. This extruder was mounted to a Kuka robot arm and water-soluble 3D parts were built using the system.
- Enhancing 3D printing with in-situ scanning – Students created a 3D scanner using a line laser and USB camera and integrated it into a Solidoodle 3D printer. They demonstrated its capabilities by scanning a part on the print bed, then using the scan information to register a subsequent print onto the part's top surface.
- Printing of carbon fiber composites using photo-curable binders – Students designed and built a machine that extrudes a continuous carbon fiber tow into a flow of resin. The machine operated by laying the resin-infused fiber into a planar trajectory, and curing the resin using a UV light source.
- Embedding electronics into additively manufactured flexible substrates – Students used a custom multi-material polymer printer, located in a research lab at the MIT Computer Science and Artificial Intelligence Laboratory to prototype concepts for embedded electronics in printed flexible substrates. A printed bracelet including an electronic identification tag was one of several objects built as final demonstrations.

The soft serve ice cream printer shown in Figure 9 was built by students Kristine Bunker (Mechanical Engineering 4$^{th}$ year undergraduate), Kyle Hounsell (Electrical Engineering Master of Engineering Student), and David Donghyun Kim (Mechanical Engineering Master of Science student). Their aim was to use additive manufacturing of ice cream to, as they noted in the project proposal, "interest children in technology and (in) pursuing science and engineering in the future". The machine comprised a freezer, a soft serve ice cream mixer, consumer-grade 3D printer, and a liquid nitrogen delivery system. Ice cream produced by the soft serve machine atop the freezer was pumped into a heated tube that entered the freezer and terminated in in close proximity to the build platform. Liquid nitrogen was sprayed at the nozzle to freeze the ice cream upon release. A three-axis moving platform, created from a modified gantry printer, allowed the generation of custom shapes. To determine the feasibility of their concept, the team measured the pressure-flow characteristics of soft serve ice cream at different temperatures. After combining each relevant module the team was able to demonstrate their concept by printing star-shaped ice cream treats. The project received widespread attention from online editorials including Tech Crunch and The Guardian [29], [30], and the team is continuing the project with the goal of licensing or commercializing the technology.

Another team developed a gantry-based machine (Figure 10) that enabled printing of optically transparent glass. The team members were graduate students from Mechanical Engineering (Michael Stern and Shreya Dave) and Media Arts and Sciences (Markus Kayser and John Klein). The machine frame was built from T-slot extrusion and housed a 3-axis positioning system and two kilns. A heating kiln mounted on the XY gantry was used to maintain the molten state of the glass as it flowed onto a ceramic build platform, which was encased in another kiln. The balance between gravity and the high viscosity of molten glass enabled controlled deposition of molten glass roads in a manner similar to FDM. The students characterized the flow rate of molten glass as a function of temperature and nozzle diameter, and used these findings to determine the gantry speed. As a demonstration of their work, the students created many artistic structures from optically transparent glass. The project was a winner of the 2015 MIT de Florez design competition, a prestigious award to recognize student achievement. The team has continued to develop the glass printing technology since the class concluded, and their work has been featured by Forbes, Popular Mechanics, and many media outlets [31], [32]. Significant improvements in part quality have been enabled by a custom nozzle geometry, and control of the surrounding temperature to maintain the desired shape after each layer is deposited.

The concept for high-throughput polymer 3D printing (Figure 11) was built by Josh Siegel, Pranay Jain, Dylan Erb, graduate students in Mechanical Engineering, along with Issac Ehrenberg, a postdoctoral associate in Mechanical Engineering. Their process, called local viscosity control (LVC), heats selective areas of a mesh wire grid to melt polymer feedstock in those areas. After local melting, the polymer is brought into contact with the build platform (or the top surface of the prior layer), and a small amount of polymer is transferred as the mesh is drawn upward and the mesh is cooled. The students demonstrated this concept by developing a single-cell prototype and control electronics, and used these to build a small polymer tower. Work continued after the semester to prototype a large grid array and develop materials enabling improved control of the temperature-viscosity transition.

*3.6 Student Feedback*

In Spring 2014, approximately 30 students were enrolled in the on-campus course. The majority of the students were from mechanical engineering but approximately one third of the students came from other departments including electrical engineering, materials science, civil engineering, aerospace engineering, and the media arts and sciences. Students from that semester reflected positively on the introductory materials and two design assignments. Some students expressed concern with the time demands and scope of the term projects, and felt the open-ended nature was more appropriate for the research environment. Another student recommended that more structure be placed in the initial organization of teams such that participants were balanced by skillset rather than interest area. We will consider the comments above in future implementations of the course. Last, we thought that a more uniform understanding of the mechanical design and key components of AM machines would have been useful to the students early in the semester, so in alternate formats of the course (described below) we have added an introductory 'Machine Anatomy' lab. In this lab, we describe the design of gantry-based and mirror-scanning AM systems, and illustrate their performance capabilities and limits. We accompany this discussion with several disassembled desktop FDM and SLA machines that the students can inspect closely.

## 4. Other Formats of the AM Course

We have also developed and taught two other formats of the AM course: one course within the MIT Master of Engineering in Manufacturing (MEngM) program [33]; as well as an industry-focused, five-day short course administered via MIT Professional Education Short Programs [34]. The MEngM course was first offered in Fall 2014 and the Short Programs course was first offered in July 2014.

The MEngM program provides graduates with a blend of theory and practice with majority focus on manufacturing processes, process control, factory system and supply chains, and business fundamentals. The MEngM AM course differed from the spring course in a number of ways. First, the course was designed as a half-time (i.e., one meeting per week throughout the semester, and lesser corresponding workload) course to meet the schedule constraints of the MEngM program. Second, the course focused more detail on the quantitative process attributes, design methods, and cost and business analysis aspects of AM. Third, the class size was smaller at 12-15 students, allowing the instructors to provide more guidance to project teams. The course still began with introductory process lectures, labs, and individual printing assignment but the bridge challenge and spring term project were substituted for an in-depth process study. Teams of students were asked perform a quantitative analysis of the attributes limiting the rate and quality of a selected process (FDM, SLM, SLA), and develop a framework for predicting part cost and compare their findings to price quotes from commercial AM suppliers. Their studies were summarized via an in-class presentation and written report due at the end of term.

The MIT Professional Education short program, titled "Additive Manufacturing: From 3D Printing to the Factory Floor", condensed the course into a single week and was tailored to fit the interests of industry participants. Because of the single week time constraint and enrollment of approximately 45 participants, the lab was split into two parallel sessions that alternated with lectures. Another challenge was balancing the focus of lecture topics during the class because of the diverse technical background of the participants. An example cohort included: technical managers and executives; design engineers in materials, aerospace, and medical industries; management and financial consultants; and independent business people. Therefore, we aimed to provide a solid foundation on the concepts of AM to allow fledgling users to understand the technology, yet include significant depth to balance the interests of expert users. In addition to special topics and applications lectures, participants completed a miniature case study project, focusing on a topic of common interest. Groups and topics were formed by the first day and teams were allowed breakout sessions to discuss ideas throughout the course. On the final day, each team made a brief presentation about their case study. We kept in touch with the participants and learned that several of the case studies resulted in continued collaboration beyond the course, including patent filings, academic research, and entrepreneurial activity.

# 5. Conclusions

To conclude, we make several recommendations for future implementations of a project-based AM course. First, we learned the importance of clarifying project goals early in the semester, to control the scope of work while maintaining high ambitions. Second, we believe a project-based AM course presents a good opportunity to seed collaborative research, and to secure financial support both during and after the term. Use of digitally archived content, both related to AM, and on supporting topics (e.g., machine design, controls, materials processing) can help adapt students of diverse academic backgrounds to the breadth and depth of the AM curriculum. This approach can leverage now well-established content delivery platforms including edX and Coursera. Third, we recognize the importance of the full suite of AM technologies (such as SLS and material jetting) and hope to have the equipment to demonstrate these in the future. Last, with reference to the industry-focused short program, we see that the background knowledge of participants in AM is growing, and the interests of participants are becoming much more specialized according to their industry and/or job function. Therefore, separate curricula can be tailored for example to introduce AM to new audiences, and to teach in-depth understanding for specific sectors such as AM of high-performance materials. Also, one of the greatest open needs for both students and practicing engineers is for education in design methods for AM, which are essential to enable us to climb the learning curve of AM and realize its true flexibility and complexity at an accessible cost.

Overall, we found it challenging to balance both breadth and depth throughout the course, because cutting-edge AM involves many topics including materials science, machine design, fluid mechanics, heat transfer, imaging, and computation. Especially in the project, teamwork among students with complementary backgrounds was highly correlated to the sophistication and novelty of their prototypes. We therefore believe that, while rooted in mechanical engineering and materials science, AM education is truly multidisciplinary. As a result, educational programs to AM processes and applications should embrace this context. We hope that our approach can be useful to others designing AM education programs to teach hands-on practice with current methods and to catalyze innovation in both AM processes and their applications.

# 6. Acknowledgements

Funding for the laboratory materials, equipment, and project expenses was kindly provided by the MIT Department of Mechanical Engineering and a faculty gift award from 3M Corporation. We thank the MIT-SUTD International Design Centre (IDC) for providing space to host the laboratory setup of the course during Spring 2014. We appreciate the help of Pierce Hayward of the MIT Department of Mechanical Engineering for use of the Instron equipment used in the bridge challenge, Charles Guan for supervising students using the IDC machine shop, Prof. David Hardt for guidance during the design and administration of the course, A.J. Perez for contributing to the course content during Spring 2014, and Profs. David Hardt and Wojciech Matusik for contributing guest lectures. We also thank the students for their enthusiasm and participation in the course during Spring 2014: Kristine Bunker, Shreya Dave, Claudio Di Leo, Jorge Duro-Royo, Isaac Ehrenberg, Dylan Erb, Kyle Hounsell, Holly Jamerson, Benjamin Jenett, Markus Kayser, Christopher Kelly, Karim Khalil, John Klein, Jared Laucks, Richard Li, Cynthia Lin, Yang Liu, Laia Mogas-Soldevila, Larissa Neitner, Scott Nill, Sumant Raykar, Joshua Sibblies, Michael Stern, Katarina Struckmann, Ramya Swamy, Louise van den Heuvel,

Katherine Luginbuhl, John Marsh, David Kim, Josh Siegel. Finally, the authors' time to prepare this manuscript was supported by a research grant from Lockheed Martin Corporation, managed by Dr. Padraig Moloney.

| Week | Day | Lecture Topic | Lab Topic | Assignment Due |
|---|---|---|---|---|
| 1 | T | Introduction to AM | Orientation | |
|  | R | Intro continued; Project Framework | | |
| 2 | T | Extrusion Methods 1 | FDM | Team and Topic Choice |
|  | R | Extrusion Methods 2 | | |
| 3 | T | Holiday (President's Day) | SLA | |
|  | R | Photopolymerization Methods 1 | | |
| 4 | T | Photopolymerization Methods 2 | 3D Scanning | |
|  | R | 3D Scanning & Metrology | | |
| 5 | T | Powder Bed Methods 1 | Open Lab | Individual Print and Analysis |
|  | R | Powder Bed Methods 2 | | |
| 6 | T | Team Meetings with Instructors | | Project Proposal |
|  | R | Team Meetings with Instructors | | |
| 7 | T | Jetting Methods | | |
|  | R | Bridge Contest | | |
| 8 | T | Holiday (Spring Break) | | |
|  | R | Holiday (Spring Break) | | |
| 9 | T | Project Design Reviews | | Design Review |
|  | R | Project Design Reviews | | |
| 10 | T | Design Review Feedback | | |
|  | R | Design for Additive Manufacturing | | |
| 11 | T | Micro/Nano AM | | |
|  | R | Bioprinting and Biocomposites | | |
| 12 | T | Holiday (Patriot's Day) | | |
|  | R | Controls | | |
| 13 | T | Digital Assembly | | Provisional Patent Draft |
|  | R | Entrepreneurship | | |
| 14 | T | Computational Design | | |
|  | R | Economics and Mfg systems | | |
| 15 | T | Project Expo | No Lab | Prototype and Report |
|  | R | Wrap-up | | |

**Figure 1**. Schedule of the additive manufacturing graduate course based on the course held at MIT in Spring 2014.

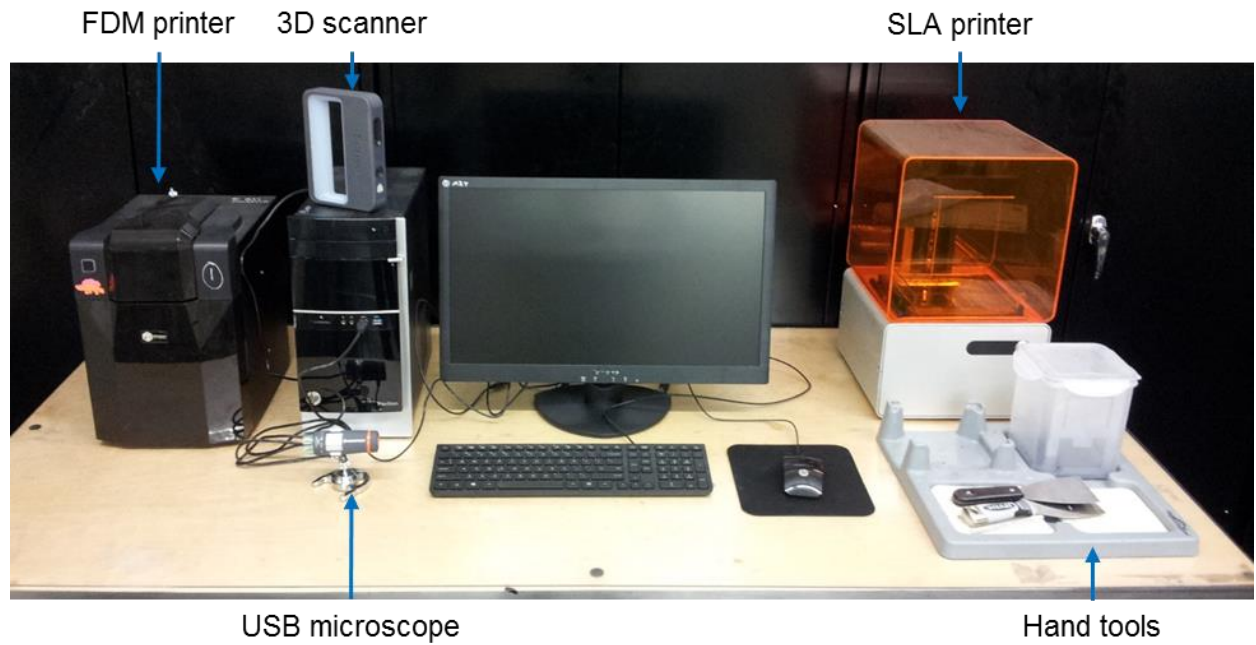

**Figure 2.** Standard lab station comprising a computer, FDM printer (Up! Mini), SLA printer (Formlabs Form1+) with post-processing setup, 3D scanner (Cubify Sense), USB microscope (Celestron 44302) and hand tools to assist with part removal and cleaning.

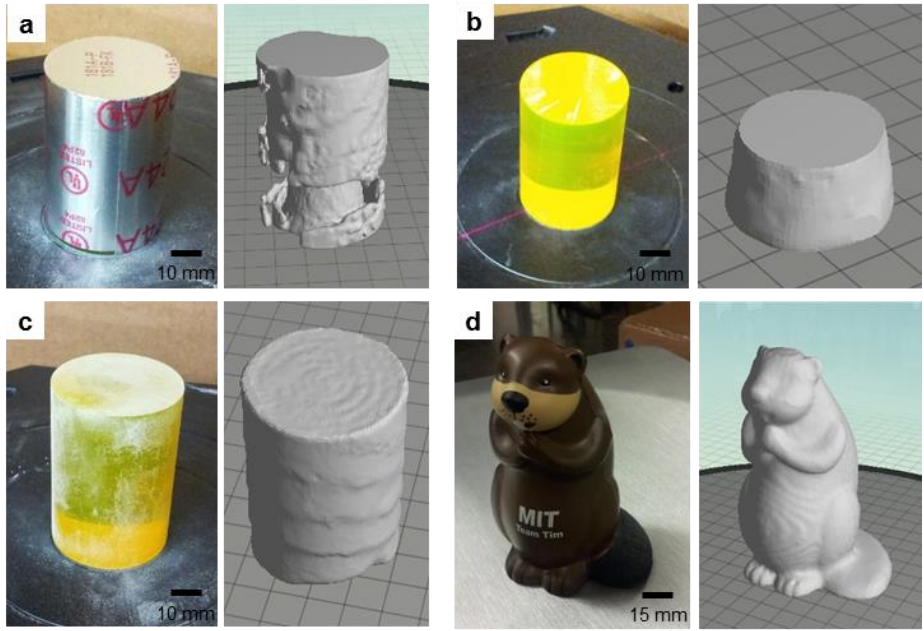

**Figure 3.** Example 3D scans of a cylindrical object illustrate defects in the scan due to: (a) reflective surfaces; (b) transparency that are in part alleviated by (c) applying a matte surface treatment. Example (d) of successful scanning a matte object (MIT mascot figurine) having moderate curvature.

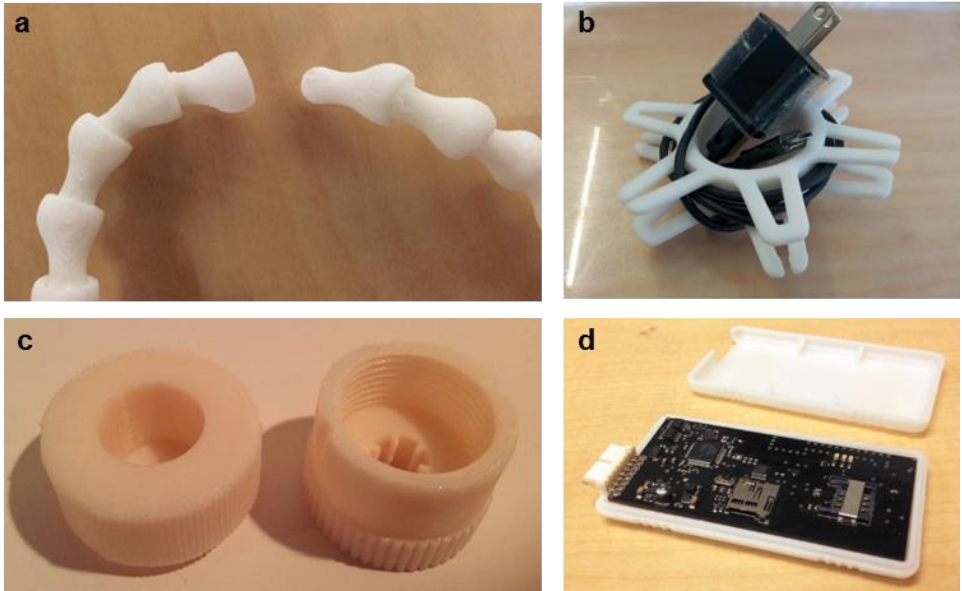

**Figure 4.** Example student submissions from the individual printing assignment made using FDM of ABS plastic: (a) flexible hose; (b) cable wrap for phone charger; (c) caps with mating threads; (d) case for printed circuit board.

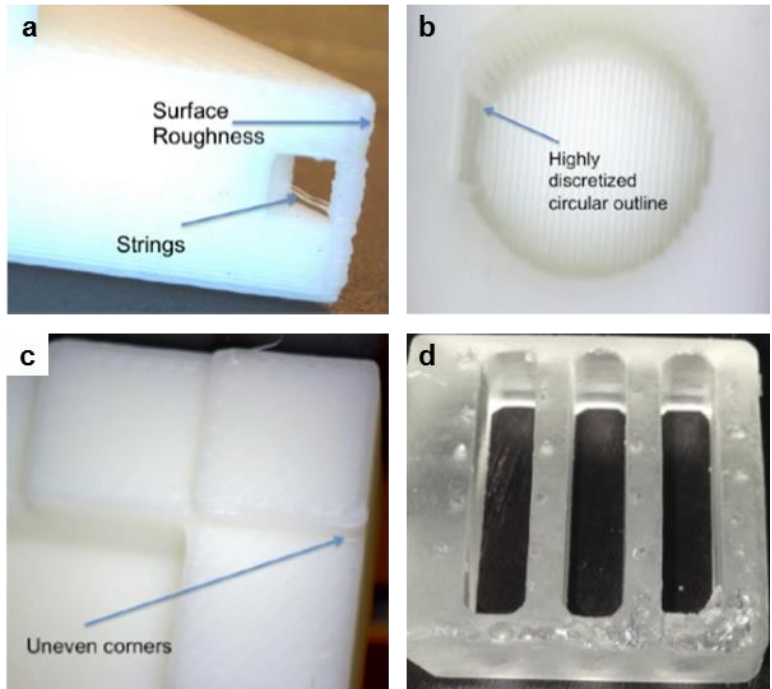

**Figure 5.** Part defects identified in student submissions to the individual printing assignment: (a) surface roughness due to layer discretization and misalignment (FDM); (b) discretization of round features due to layer thickness (FDM); (c) bulging due to extrusion of excess material at part corners (FDM); (d) surface roughness from breakaway support structures (SLA). The annotations are as submitted by the students.

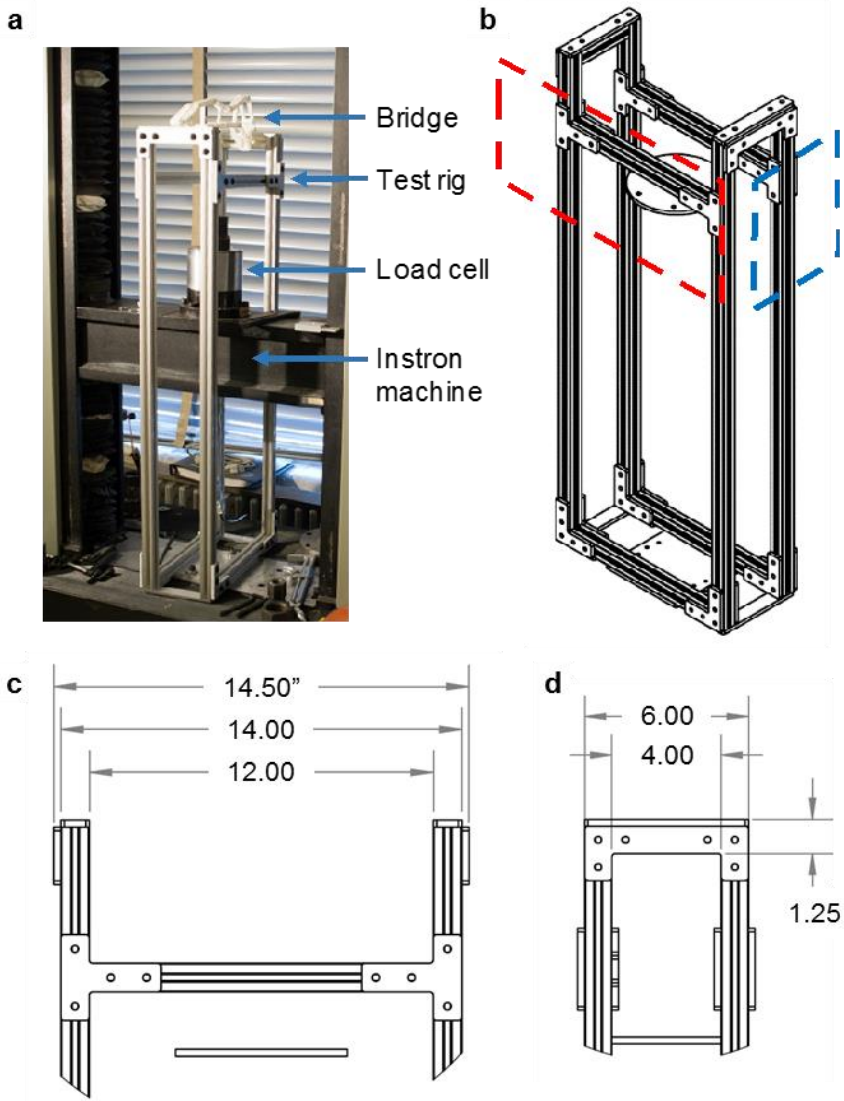

**Figure 6**. Bridge challenge test setup: (a) rig mounted to Instron machine, prepared to test a bridge by applying a vertical load downward at the center of the span; (b-d) drawings of the rig with dimensions in inches.

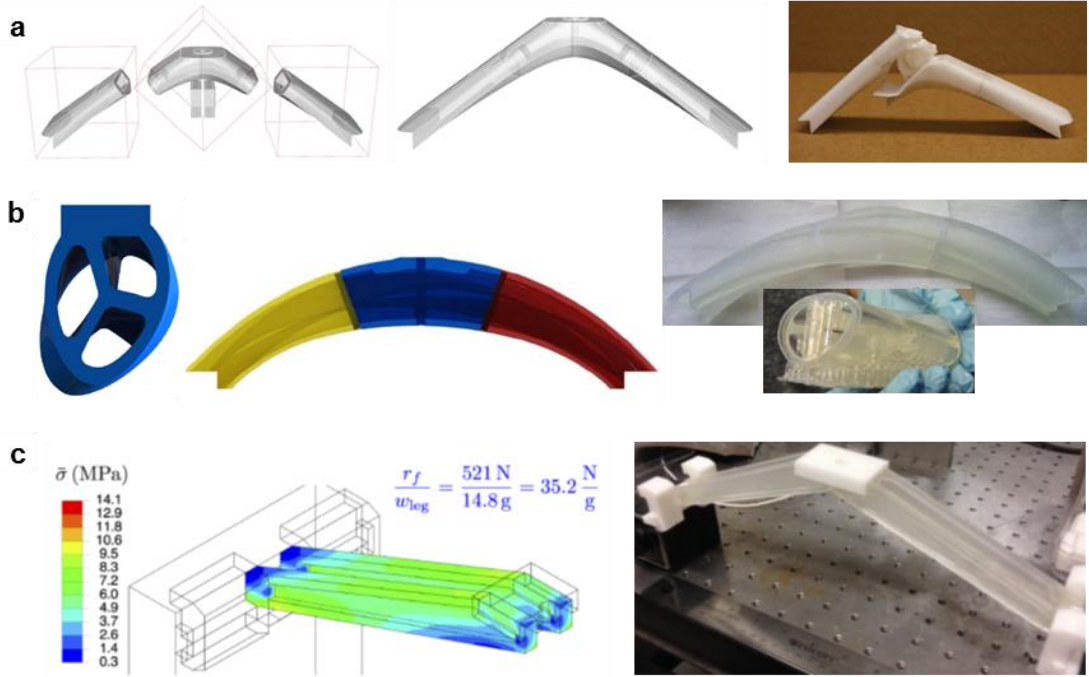

**Figure 7.** Selected student team bridge designs included: (a) multi-piece bridge made with topology optimization (b) bio-inspired design; (c) linkage design and use of finite element analysis (FEA) to simulate load-displacement behavior.

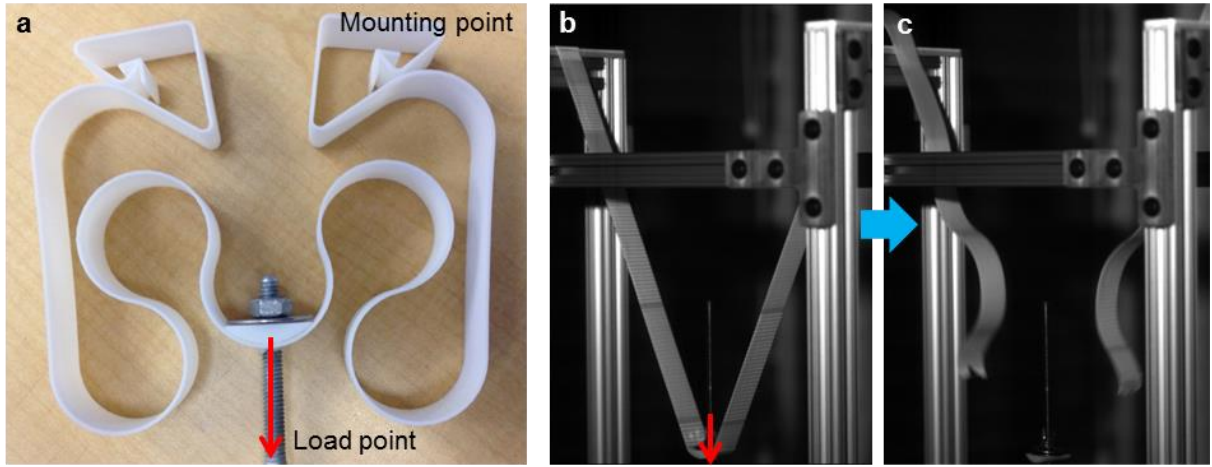

**Figure 8.** The winning bridge design: (a) as-printed bridge, which was a single strip of ABS with layers parallel to the image plane and integral load and mounting features; (b) bridge stretched during early stage of test; (c) bridge immediately after failure during test. The bridge weighed 24 grams and supported 1014 N of load at the moment of failure. Two identical bridges were tested. One failed at the load point, and another failed along the span, at what appeared to be an interlayer defect arising from the FDM process.

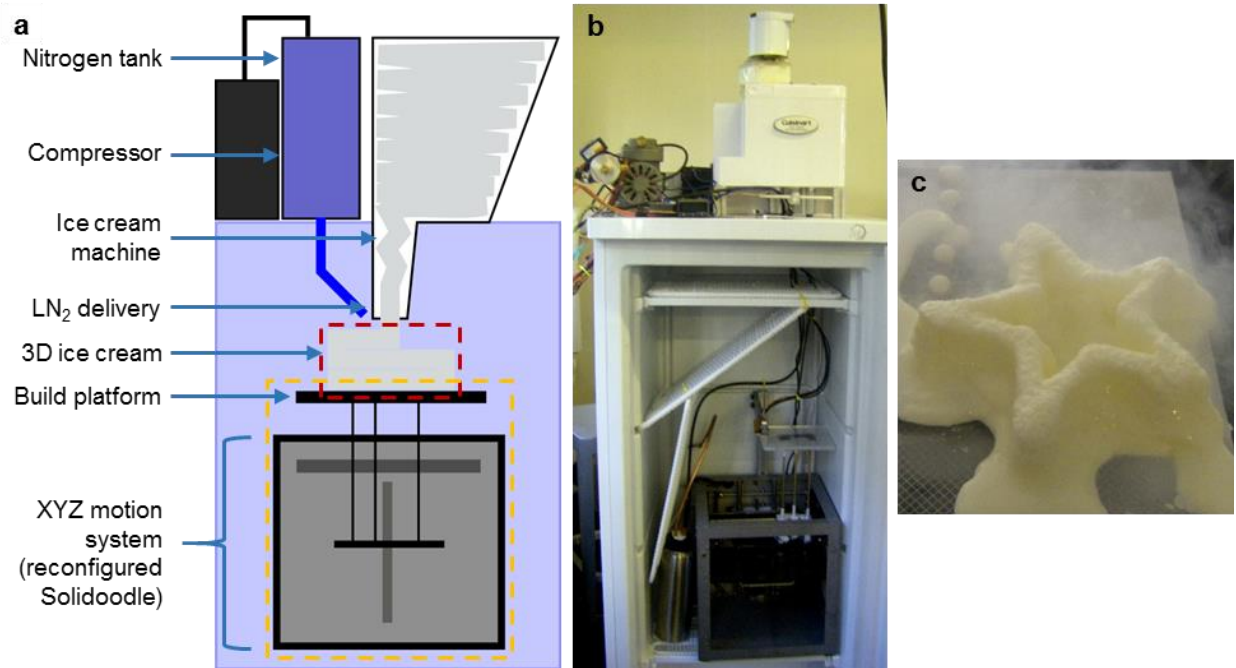

**Figure 9.** The ice cream 3D printer: (a) system schematic and major components; (b) overall machine with the freezer door open; (c) soft-serve ice cream printed into a star shape.

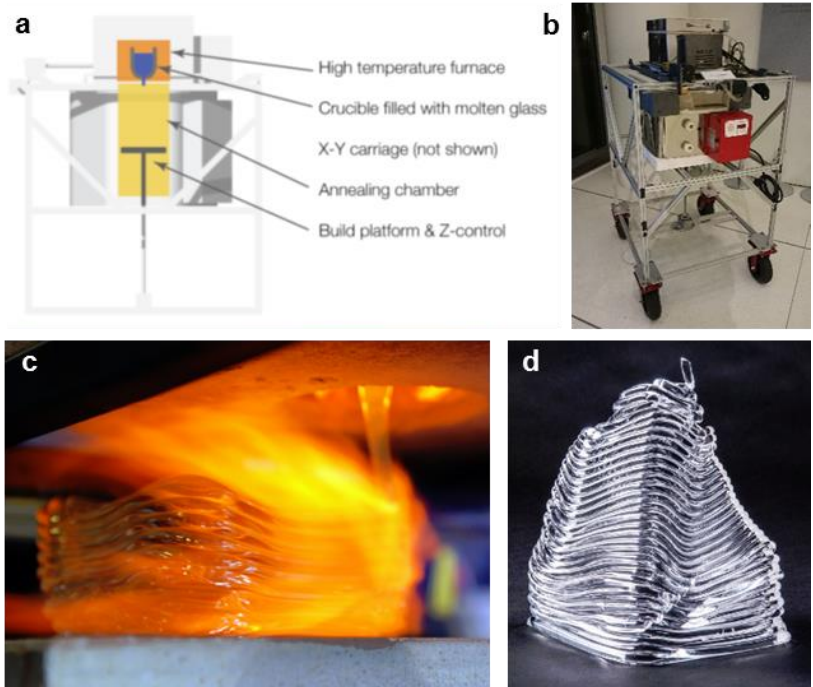

**Figure 10**. The glass printer: (a) schematic of first-generation machine built during the AM class; (b) second-generation machine on display in the lobby of the MIT Media Lab; (c) image taken during glass printing by the first-generation machine; (d) printed prototype glass object that was displayed at the class expo.

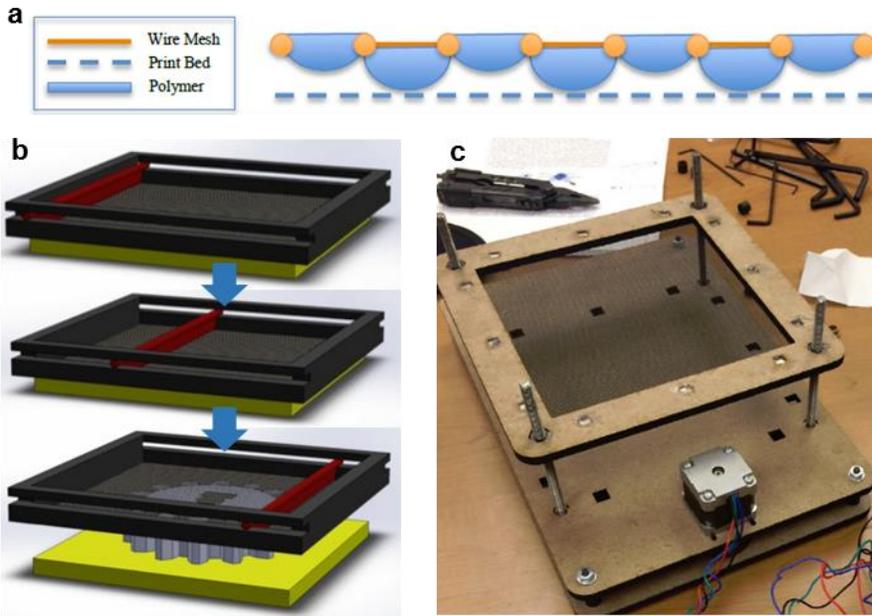

**Figure 11**. Local viscosity control (LVC) printing: (a) mechanism of printing by selectively heating sections of a wire mesh; (b) proposed method of depositing polymer droplets in parallel across the build area; (c) prototype mechanism under construction for the class project.